# Maximization and Minimization of Interfacial Thermal Conductance by Modulating the Mass Distribution of Interlayer

Lina Yang[1#], Xiao Wan[2#], Dengke Ma[3], Yi Jiang[1*], Nuo Yang[2*]

[1]School of Aerospace Engineering, Beijing Institute of Technology, Beijing 100081, China

[2]State Key Laboratory of Coal Combustion, and School of Energy and Power Engineering, Huazhong University of Science and Technology, Wuhan 430074, China

[3]NNU-SULI Thermal Energy Research Center (NSTER) and Center for Quantum Transport and Thermal Energy Science (CQTES), School of Physics and Technology, Nanjing Normal University, Nanjing, 210023, China

Email: jyjybit@bit.edu.cn (Y. Jiang), nuo@hust.edu.cn (N. Yang)

## Abstract

Tuning interfacial thermal conductance has been a key task for the thermal management of nanoelectronic devices. Here, it is studied how the interfacial thermal conductance is great influenced by modulating the mass distribution of the interlayer of one-dimensional atomic chain. By nonequilibrium Green's function and machine learning algorithm, the maximum/minimum value of thermal conductance and its corresponding mass distribution are calculated. Interestingly, the mass distribution corresponding to the maximum thermal conductance is not a simple function, such as linear and exponential distribution predicted in previous works, it is similar to a sinusoidal curve around linear distribution for larger thickness interlayer. Further, the mechanism of the abnormal results is explained by analyzing the phonon transmission spectra and density of states. The work provides deep insight into optimizing and designing interfacial thermal conductance by modulating mass distribution of interlayer atoms.

# 1. Introduction

As the developing of nanotechnology, the dimensions of materials shrink to nanoscale, and the interfacial thermal resistance plays a key role in determining the overall thermal transport properties of nanostructured devices[1-7]. The limitation of heat dissipation due to interfacial thermal conductance (ITC) has been a bottleneck for nanoelectronics. At present, enhancing interfacial thermal conductance is a critical task for researchers. On the other hand, reducing interfacial thermal conductance is also required for thermal insulators and thermoelectrics[8,9].

So far, numerous studies have been done to enhance interfacial thermal conductance. The roughness of interface (triangular-shaped interfaces) can enhance the ITC by a factor greater than three[10]. Nanopillar arrays patterned interface can also help interfacial thermal transport because of the enlarged effective contact area[11]. Atomic mixing or disorder at the interface can increase the ITC due to the overlap of the density of states (DOS), which increases the phonon transmission[12-14]. Strengthening the bonds at interface can increase ITC[15-17], it was shown that the ITC reaches its maximum if the interfacial coupling is the harmonic mean of the spring constants of the contact materials.

Another way to enhance ITC is the insertion of a nanoscale thin interlayer at the interfaces[14,18-22]. It is found that a multi-atom-thick thin film at the interface can increase the ITC for a solid-solid interface, because the phonon DOS of the interlayer bridges the two different phonon DOS of the contacts by nonequilibrium Green's functions (NEGF) and nonequilibrium molecular dynamics method[18,19]. By using the impedance matching criterion, Polanco *et al*. found that the arithmetic mean for mass and the harmonic mean for spring constant in one-dimensional (1D) junction model can largely increase the ITC[20]. Later, they found the maximum ITC happens when the interlayer mass is close to the geometric mean of the contact masses for three-dimensional crystal interfaces[19].

To further increase the ITC, graded mass distribution of interlayer is investigated. Recently, Zhou *et al*. reported that the ITC of interface with interlayer of linear mass distribution can be enhanced by 6-fold compared with that of abrupt interface by nonequilibrium molecular dynamics method in the Ar-heavy and Ar crystal system[23]. Rastgarkafshgarkolaei *et al*. also studied the interface with interlayer of exponential mass distribution and found that the ITC initially increases as the number of layers increases due to the larger phonon transmission because of the improved DOS bridging, but quickly saturates[24]. Moreover, they found that the exponential mass distribution has better performance than the linear mass distribution in enhancing the ITC. Later, Xiong *et al*. found that the ITC of 1D atomic junction with interlayer of both geometric graded mass and geometric graded coupling can be enhanced nearly up to 6-fold by NEGF method[25]. Recently, the machine learning algorithms, which can identify optimized structures, have been widely applied in designing interfacial structures[26-28]. Ju *et al*. studied the minimization of ITC by modulating the mass distribution of interlayer by NEGF and Bayesian optimization[27], and found that the aperiodic interlayer can lead to the smallest ITC of Si-Ge interfaces, which is obviously different from periodic interlayer. Recent work by Chowdhury *et al*. also found that the random multilayers has minimum ITC by a genetic algorithm based approach and molecular dynamics simulations[29].

In this work, the ITC of interface of 1D atomic chain is maximized/minimized by modulating the mass distribution of interlayer using nonequilibrium Green's function and machine learning algorithm. Moreover, the performance of linear, the exponential and periodic mass distribution are compared with that of mass distribution with maximum/minimum ITC. Furthermore, the phonon transmission and phonon density of states are calculated to understand the underlying physical mechanism of the mass distribution effect on the interfacial thermal conductance.

## 2. Model and method

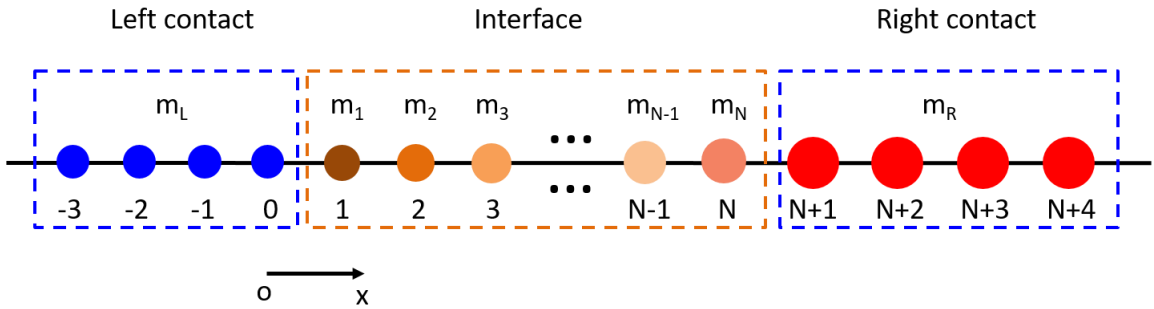

Figure 1. Schematic of the 1D atomic chain. The numbers below the atoms denote the respective atom index. The mass of the atoms in the left and right contact are $M_L$ and $M_R$, respectively. The mass of the atoms with index $i$ is $M_i=m_i\times10^{-26}$ kg. The interlayer $N$ is the number of atoms in the interlayer. The spring constant is $k$, which is invariant in the whole system. Here, we set $M_L=m_L\times10^{-26}$ kg, $M_R=m_R\times10^{-26}$ kg, $m_L=1.0$, $m_R=4.0$ and $k=4.0$ N/m, the cutoff angular frequency of left and right contact are 40 THz and 20 THz, respectively. The distance between atoms is $La=0.5$ nm. The 1D atomic chain is along $x$ direction. The origin of coordinate is at the position of atom with index 0. The position of atom in the interlayer is $La\times i$.

To maximize/minimize ITC, the mass distribution of interlayer of 1D atomic chain is investigated by nonequilibrium Green's function method[13,30-32]. The 1D atomic chain and the settings of the system are shown in Figure 1. The dashed lines divide the system into three parts: the left semi-infinite contact, interface and the right semi-infinite contact. The numbers below the atoms denote the respective atom index. The spring constant is $k$, which is invariant in the whole system. The mass of the atom in the left (right) contact is $M_L$ ($M_R$), here, we set $M_L = m_L \times 10^{-26}$ kg, $M_R = m_R \times 10^{-26}$ kg, $m_L = 1.0$, $m_R = 4.0$ and $k = 4.0$ N/m, the cutoff angular frequency of the left and the right contact are 40 THz and 20 THz, respectively. The interlayer $N$ is the number of atoms in the interlayer of the interface. The mass of the atoms in the interlayer is $M_i = m_i \times 10^{-26}$ kg, where $i$ is the index from 1 to $N$. The distance between atoms is $La = 0.5$ nm, which is invariant in the system. The 1D atomic chain is along $x$ direction. The origin of the coordinate is at the position of atom with index 0. The position of atom in the interlayer is $La \times i$. The interfacial thermal conductance is studied at temperature of 100 K.

The equation of motion of the system can be written by the following matrix form:

$$\left( \omega^2 \mathbf{I} - \mathbf{H} \right) \boldsymbol{\psi} = 0 \tag{1}$$

where $\omega$ is the angular frequency, $\mathbf{I}$ is the identity matrix, $\boldsymbol{\psi}$ is the eigenvectors of the whole system, and $\mathbf{H}$ is the dynamical matrix representing the atomic interactions. In this work, the $\mathbf{H}$ is in the following form:

$$\mathbf{H} = \left\{ H_{pq} \right\} = \frac{1}{\sqrt{m_p m_q}} \left\{ \begin{array}{ll} \dfrac{\partial^2 U}{\partial u_p \partial u_q} & if \ p \neq q \\ -\sum\limits_{j \neq q} \dfrac{\partial^2 U}{\partial u_q \partial u_j} & if \ \mathrm{p=q} \end{array} \right\} \tag{2}$$

where $U$ is the interatomic potential energy function and $u$ is the spatial displacement of atom. In this work, the spring constant $k$ is invariant in the system, therefore, $\dfrac{\partial^2 U}{\partial u_p \partial u_q} = k$ , if $p \neq q$ . As the system is divided into three parts, the dynamical matrix also can be written into the following form:

$$\mathbf{H} = \begin{bmatrix} \mathbf{H}_L & \mathbf{H}_{LD} & 0 \\ \mathbf{H}_{DL} & \mathbf{H}_C & \mathbf{H}_{DR} \\ 0 & \mathbf{H}_{RD} & \mathbf{H}_R \end{bmatrix} \tag{3}$$

where $\mathbf{H}_C$ is the dynamical matrix of the interface, $\mathbf{H}_{LD}$ ($\mathbf{H}_{RD}$) is the coupling matrix between the left (right) contact and the interface. The interface retarded Green's function is $\mathbf{G}_D^{ret}(\omega) = \left(\omega^2 \mathbf{I} - \mathbf{H}_D - \mathbf{\Sigma}_L - \mathbf{\Sigma}_R\right)^{-1}$, $\mathbf{\Sigma}$ are self-energies which are defined as $\mathbf{\Sigma}_L = \mathbf{H}_{DL} \mathbf{g}_L^{00} \mathbf{H}_{LD}$. The left and right uncoupling retarded Green's functions are $\mathbf{g}_L^{00} = \left[\left(\omega^2 + \eta i\right)\mathbf{I} - \mathbf{H}_L\right]^{-1}$ and $\mathbf{g}_R^{00} = \left[\left(\omega^2 + \eta i\right)\mathbf{I} - \mathbf{H}_R\right]^{-1}$.[33] The phonon transmission coefficient is calculated as:

$$\Xi(\omega) = Tr\left(\mathbf{\Gamma}_R \mathbf{G}_D^{ret} \mathbf{\Gamma}_L \mathbf{G}_D^{ret\dagger}\right) \tag{4}$$

where $\mathbf{\Gamma}_L = i\left(\mathbf{\Sigma}_L - \mathbf{\Sigma}_L^\dagger\right)$ and $\mathbf{\Gamma}_R = i\left(\mathbf{\Sigma}_R - \mathbf{\Sigma}_R^\dagger\right)$. Finally, the temperature dependent thermal conductance of the interface is calculated by the Landauer formalism[34] :

$$\sigma(T) = \frac{1}{2\pi} \int_0^{\omega_{max}} \hbar \omega \frac{\partial n(\omega, T)}{\partial T} \Xi(\omega) d\omega \tag{5}$$

where $n(\omega, T)$ is Bose-Einstein distribution, $T$ is temperature.

In machine learning part of minimization of ITC, four basic elements are required when conducting Bayesian optimization: the descriptor, evaluator, calculator, and optimization method. In this study, $m_i$ of atoms in the interlayer is used as descriptors to represent the possible interfacial structures in the minimization. For the evaluator, the value of ITC is chosen to quantitatively evaluate the performance of each configuration. We employ open-source Bayesian optimization library COMBO to perform the minimization process,[35] which has been applied in recent work[26]. The details of Bayesian optimization can be found in the Supplemental Material.

### 3. Results

In order to maximize the ITC by modulating the mass distribution, the mass of atoms in the interlayer is varied in the range from $M_L$ to $M_R$. Here, the mass interval is set as $0.15 \times 10^{-26}$ kg. Figure 2 (a) shows the schematic mass distribution of interlayer of 1D atomic chain with the maximum ITC, and Figure 2 (b) shows the mass distribution with maximum ITC versus the normalized position of atoms. The interlayer $N$ varies from 1 to 8. When there is only one atom in the interlayer, linear mass distribution ($M_I = 2.5 \times 10^{-26}$ kg, arithmetic mean of the masses of the contacts), tends to exhibit the maximum ITC. On the contrary, as interlayer $N$ increases to 7 and 8, the mass distribution with the maximum ITC is close to a nonlinear distribution, which is neither linear nor exponential distribution predicted by previous work[19,24,25], is similar to a sinusoidal curve around linear distribution for larger thickness interlayer. In addition, the intersection point between mass distribution with maximum ITC and linear distribution is closer to the contact that has large value of atom mass. The maximum ITC mass distribution at room temperature is the same as that in Figure 2 (b) at 100 K, the calculation and results are shown in the Supplemental Material.

Further, the mass distribution with minimum ITC is investigated. Figure 2 (c) shows the schematic mass distribution of interlayer with minimum ITC, and Figure 2 (d) shows the mass distribution with minimum ITC versus normalized position of atoms. It is found that the mass of atom in the interlayer is either $M_L$ or $M_R$ to achive the minmum ITC. The results also demonstrate that aperiodic superlattices interlayer can have lower thermal conductance than periodic superlattice interlayer, which is consistent with the results of minimum ITC of Si-Ge interface by Bayesian optimization[27]. The minimum ITC mass distribution at room temperature is the same as that in Figure 2 (d) at 100 K, the calculation and results are shown in the Supplemental Material.

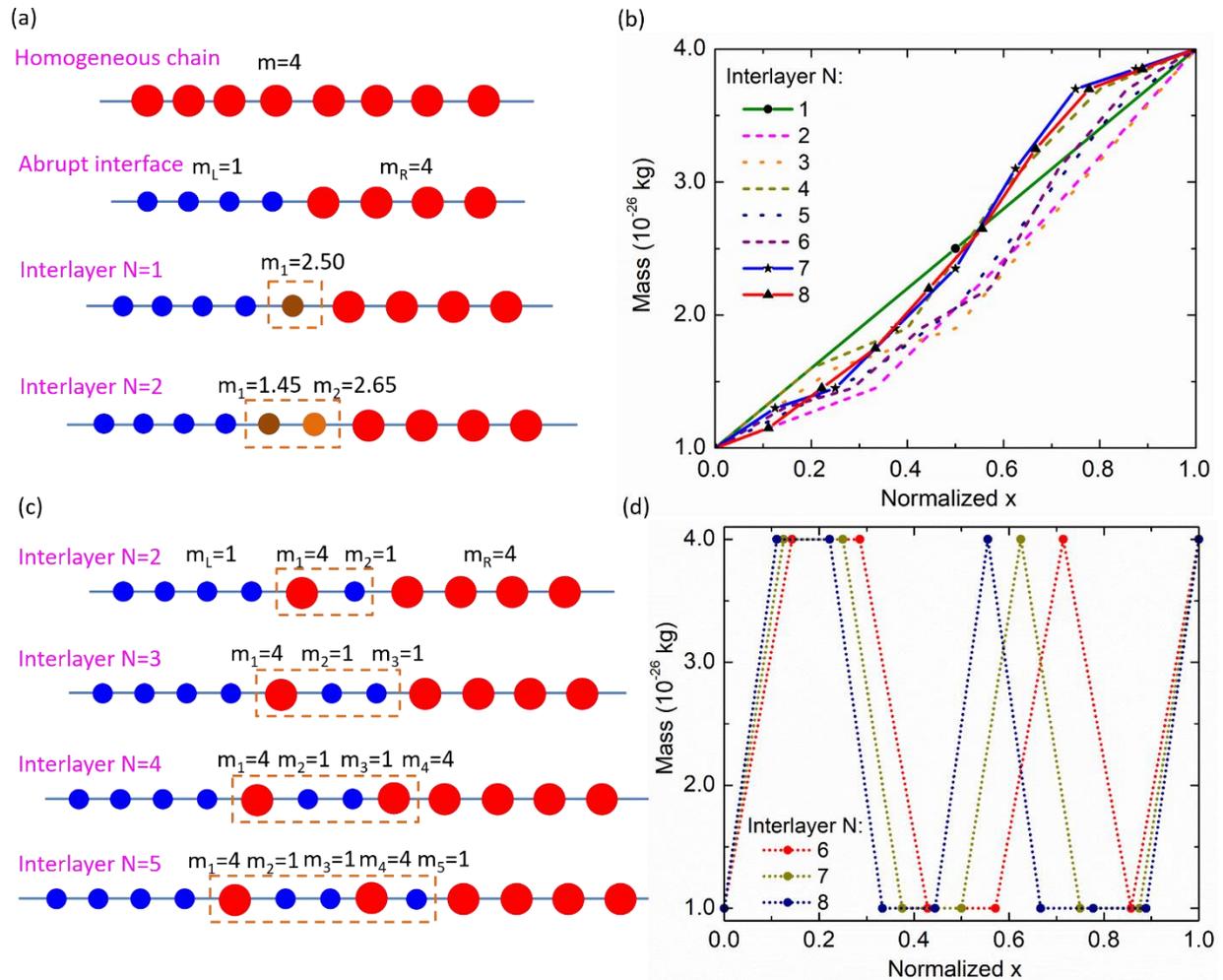

Figure 2. (a) Schematic mass distribution of interlayer in 1D atomic chain with maximum ITC. The homogeneous chain and the abrupt interface are shown for comparison. (b) The mass distribution with maximum ITC versus normalized *x*. (c) Schematic mass distribution of interlayer in 1D atomic chain with minimum ITC. (d) The mass distribution with minimum ITC versus normalized *x*. The number of atoms in the interlayer (interlayer *N*) is varied from 1 to 8.

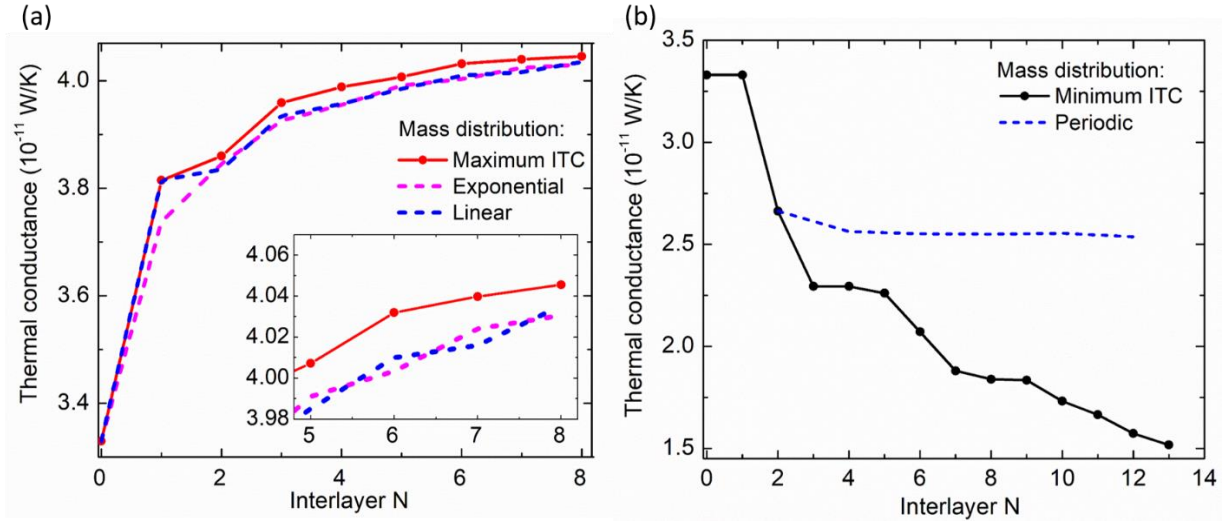

Figure 3. Interfacial thermal conductance versus the number of atoms in the interlayer (interlayer *N*). (a) Interfacial thermal conductance of interface with maximum ITC (red solid line with red dots), exponential (pink dashed line) and linear (blue dashed line) mass distribution versus interlayer *N*. The inset shows a zoom-in of the region around $4 \times 10^{-11}$ W/K. (b) Interfacial thermal conductance of interface with minimum ITC (black solid line with black dots) and periodic (blue dashed line) mass distribution. The periodic length is $2 \times La$. For *N* from 9 to 13, the mass distributions with minimum ITC are calculated by Bayesian optimization method. The temperature is set as 100 K.

Furthermore, we study the changes of maximum and minimum ITC as the increase of the number of atoms in the interlayer (interlayer *N*), and the results are shown in Figure 3. The temperature is set as 100 K. For comparison, we also calculate the ITC of interlayer with linear, exponential and periodic mass distribution. It was found that the exponential mass distribution has better performance than the linear mass distribution in 3D cubic crystal structure[24]. However, neither the linear mass distribution nor the exponential mass distribution can achieve the maximum ITC as the increase of interlayer $N$ shown in Figure 3 (a). When the interlayer $N$ is larger than 1, the linear and exponential mass distribution have similar value of ITC, but which one has better performance depends on interlayer $N$ as shown in the insets in Figure 3 (a). To be emphasized, only harmonic interaction is considered in this work. Previous study found that anharmonic interaction can significantly facilitate thermal transport[19,36-38], because it help phonons thermalize to frequencies with larger transmission rate. Therefore, additional increase of the maximum ITC can be expected if the anharmonic interaction is considered.

Moreover, the minimum ITC is studied based on the machine-learning algorithm, Bayesian optimization, as the interlayer $N$ is extended to 13, which is shown in Figure 3 (b). The details of Bayesian optimization can be found in the Supplemental Material. When there is only one atom in the interlayer, the ITC is the same as that of the abrupt interface. On the whole, the minimum ITC is greatly decreased as interlayer $N$ increases, while the ITC of interface with periodic mass distribution (period length $2 \times La$) is slowly changed. Hence, the gap between the ITC of interface with these two types of mass distributions gradually increases with the increase of interlayer $N$. Overall, by tuning the mass distribution, the ITC can be modulated in a wide range.

To understand the enhancement of ITC, the phonon transmission coefficients at the interfaces are calculated. Figure 4 (a) compares the phonon transmission coefficient at interface with maximum ITC, linear and exponential mass distribution. Unexpectedly, the phonon transmission for maximum ITC is smaller than that for linear and exponential mass distribution at low frequency ($< 6$ THz), but it has obviously larger phonon transmission at high frequency ($> 18$ THz), which leads to increase of the ITC. In addition, phonon transmission gradually increases as the interlayer $N$ increases for mass distribution with maximum ITC as shown in Figure 4 (b), which causes the enhancement of the ITC in Figure 3 (a). The details of phonon transmission with the increase of interlayer N are shown in the Supplemental Material. Besides, both the minimum ITC and periodic mass distribution cause oscillating characteristic of phonon transmission as the frequency increases shown in Figure 4 (c). Moreover, the phonon transmission for minimum ITC is greatly decreased at high frequency as the increase of interlayer $N$ shown in Figure 4 (d), which leads to the reduction of ITC. Therefore, modulating the mass distribution can greatly control the phonon transmission coefficients.

For further understanding the effect of mass distribution on ITC, phonon density of states are also studied. As shown in Figure 5 (a), the DOS of interlayer with maximum ITC is larger when angular frequency is smaller than 20 THz compared with that of linear and exponential mass distribution. The analyses of the local DOS of the atom with index 8 show that the local DOS for maximum ITC is closer to the DOS of homogeneous chain compared with that for linear and exponential mass distribution shown in Figure 5 (b). Moreover, the phonon local DOS of atoms with index from 1 to 8 are flattened and gradually increased when the frequency is less than 20 THz for maximum ITC as shown in Figure (c). Therefore, Figure 5 (b) and (c) indicate that the mass distribution of the interlayer with maximum ITC can contribute a better bridging between the DOS of the contacts, which leads the enhancement of the ITC. On the contrary, the local DOS of the atoms in the interlayer with minimum ITC oscillates as frequency increases shown in Figure 5 (d). In addition, the local DOS of atoms with mass $M_L$ (atom index 3 and 8) is suppressed to low values in a wide frequency range, which cause the great reduction of the ITC.

Therefore, modulating the mass distribution of interlayer can tune the DOS towards that of homogeneous chain to maximally increase ITC or make it oscillate to minimize the ITC.

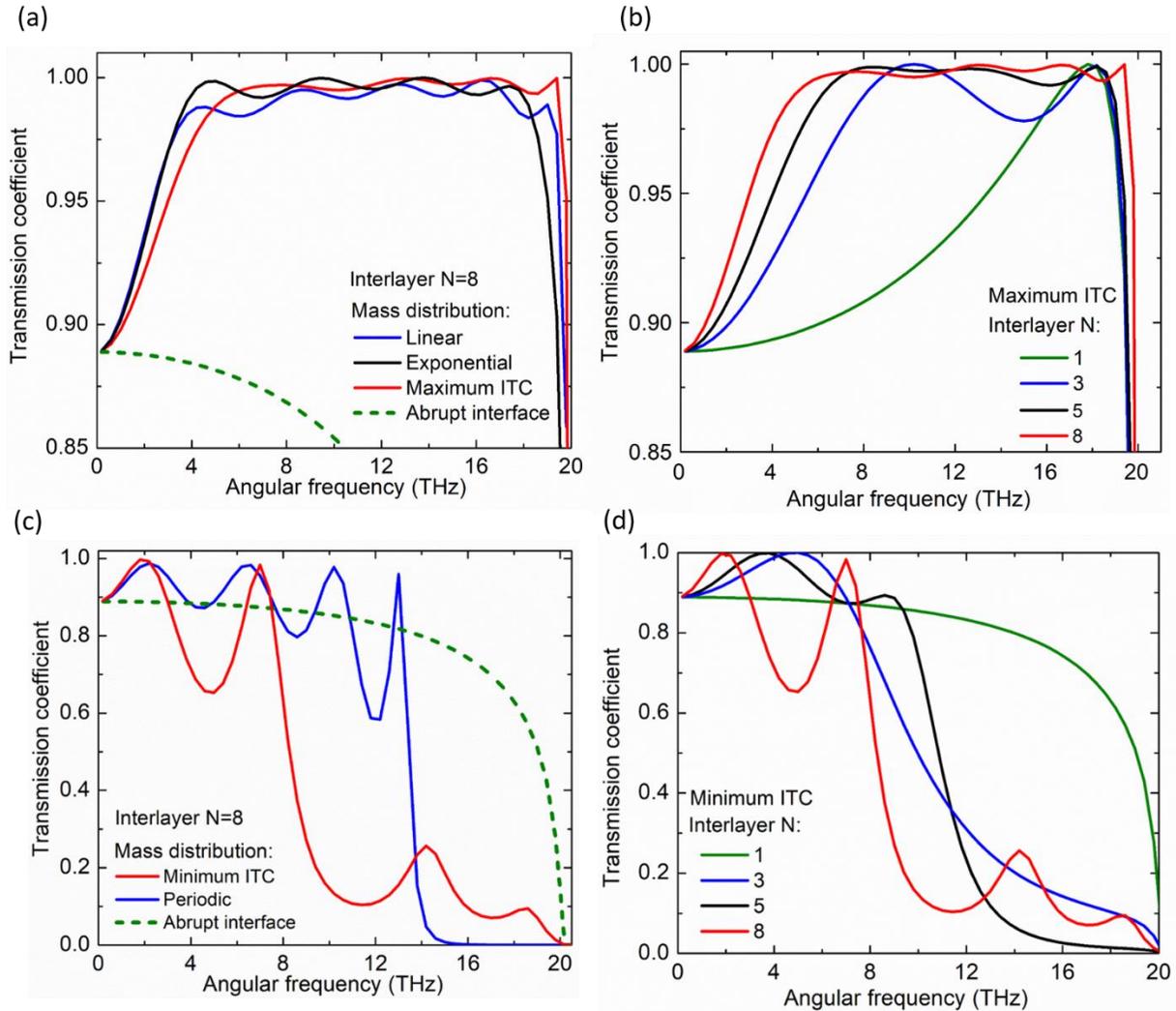

Figure 4. (a) Phonon transmission coefficients for interface with maximum ITC (red line), linear (blue line) and exponential (black line) mass distribution and for abrupt interface (green dashed line) versus angular frequency. (b) Phonon transmission coefficients for maximum ITC with interlayer $N$=1 (green line), $N$=3 (blue line), $N$=5 (black line) and $N$=8 (red line) versus angular frequency. (c) Phonon transmission coefficients for interface with minimum ITC (red line) and periodic (blue line) mass distribution and for abrupt interface (green dashed line) versus angular frequency. (b) Phonon transmission coefficients for minimum ITC with interlayer $N$=1 (green line), $N$=3 (blue line), $N$=5 (black line) and $N$=8 (red line) versus angular frequency.

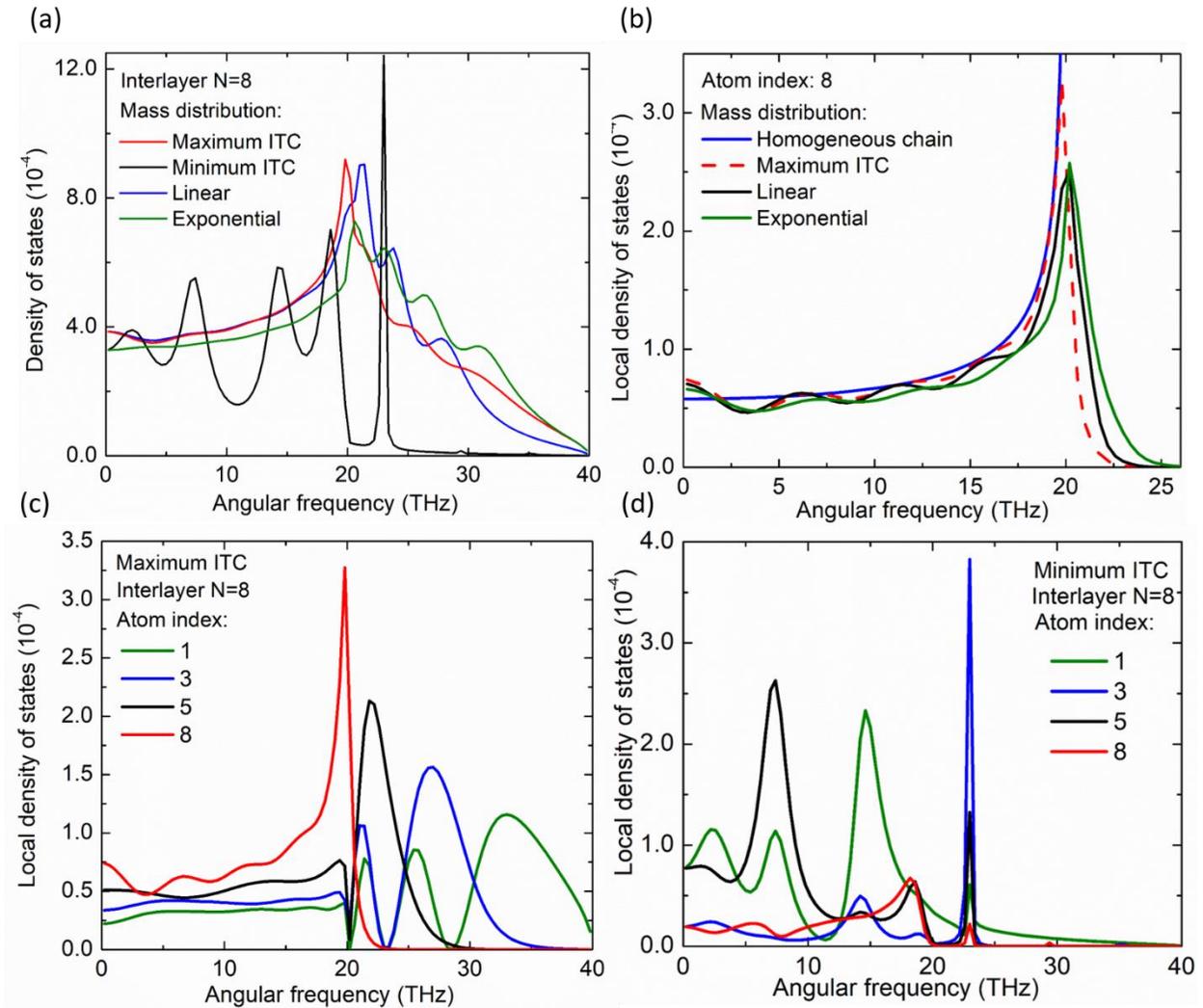

Figure 5. Phonon density of states. (a) DOS of interlayer with maximum ITC (red line), minimum ITC (black line), linear (blue line), and exponential (green line) mass distribution versus angular frequency. (b) Local DOS of atom with index 8 in the interlayer with maximum ITC (red dashed line), linear (black line) and exponential (green line) mass distribution versus angular frequency. The DOS of the homogeneous chain (blue line) with mass $4 \times 10^{-26}$ kg and $k$=4.0 N/m is also shown. The interlayer $N$ equals 8. (c) Local DOS of atoms with index 1 (green line), 3 (blue line), 5 (black line) and 8 (red line) in the interlayer with maximum ITC versus angular frequency. (d) Local DOS of atoms with index 1 (green line), 3 (blue line), 5 (black line) and 8 (red line) in the interlayer with minimum ITC versus angular frequency.

## 4. Conclusion

In this work, we studied the mass distribution of the interlayer of 1D atomic chain to achieve maximum and minimum ITC by nonequilibrium Green's function method and machine learning algorithm. We found that the mass distribution with maximum ITC is not a simple function, such as linear and exponential distribution predicted in previous works, it is similar to a sinusoidal curve around linear distribution for larger thickness interlayer. In addition, the minimum ITC is obtained by aperiodic arrangement of $M_L$ and $M_R$. The phonon transmission analyses show that the interlayer with maximum ITC has larger transmission than that of linear and exponential mass distribution at higher frequency, while interlayer with minimum ITC has phonon transmission with oscillating characteristic. Moreover, the local DOS of the interlayer with maximum ITC gradually increases as the atom index increases and is close to the DOS of the homogeneous chain, which leads to the enhancement of the ITC. On the contrary, the local DOS of minimum ITC are suppressed to low values in a wide frequency range, which leads to a great reduction of ITC. Our work provides deep insight into optimizing ITC by the way of modulating the mass distribution of interlayer and offer guidance for designing interfaces.

**Acknowledgements**

The work was sponsored by Fundamental Research Funds for the Central Universities (2019kfyRCPY045). The authors thank the National Supercomputing Center in Tianjin (NSCCTJ) and China Scientific Computing Grid (ScGrid) for providing assistance in computations.

**Data availability**

The raw/processed data required to reproduce these findings cannot be shared at this time due to technical or time limitations